\documentclass[a4paper,11pt]{article}
\pdfoutput=1

\usepackage{amsmath, amssymb, bm}
\usepackage[top=1in, bottom=1.2in, left=0.7in, right=0.7in]{geometry}
\usepackage[]{graphicx}
\usepackage[font=scriptsize,labelfont=bf]{caption}
\usepackage[caption=false]{subfig}
\usepackage{slashed}
\usepackage{textcomp}
\usepackage{authblk}
\usepackage{hyperref}
\newcommand{\be}{\begin{eqnarray}}
\newcommand{\ee}{\end{eqnarray}}

\begin{document}

\title{\Large{\bf{Do photons travel faster than gravitons?}}}
\author{Damian Ejlli}

\affil{\emph{\normalsize{Department of Physics, Novosibirsk State University, Novosibirsk 630090 Russia}}}

\date{}

\maketitle

\begin{abstract}
The vacuum polarization in an external gravitational field due to one loop electron-positron pair and one loop millicharged fermion-antifermion pair is studied. Considering the propagation of electromagnetic (EM) radiation and gravitational waves (GWs) in an expanding universe, it is shown that by taking into account QED effects in curved spacetime, the propagation velocity of photons is superluminal and can exceed that of gravitons. We apply these results to the case of the GW170817 event detected by LIGO. If the EM radiation and GWs are emitted either simultaneously or with a time difference from the same source, it is shown that the EM radiation while propagating with superluminal velocity, would be detected either in advance or in delay with respect to GW depending on the ratio of millicharged fermion relative charge to mass $\epsilon/m_\epsilon$. 
\end{abstract}

\vspace{+1cm}

\section{Introduction}
\label{sec:1}

The detection of the GWs events by LIGO \cite{Abbott:2016blz}, undoubtedly has opened a new window in the study of the early and contemporary universe. The importance of such observations not only relies on the study of GWs but also gives a unique opportunity to test unobserved quantum gravity effects. One important information that we get from the observation of these events, is that GWs or gravitons, propagate from the sources to the detector with speed\footnote{In this paper I work with the rationalized Lorentz-Heaviside natural units $c=\hbar=k_B=\varepsilon_0=\mu_0=1$ with $e^2=4\pi\alpha$ and metric with signature $\eta_{\mu\nu}=\text{diag}(1, -1, -1, -1)$.} $v_\text{gw}=1$. This fact means that there are tight constraints\footnote{Based on the fact that LIGO observations indicate that $v_\text{gw}=1$, here we do not consider the vacuum polarization effects for GWs but only for the EM radiation. } on possible modifications of GWs dispersion relation\footnote{For indirect constraints on GW dispersion relations see Ref. \cite{Kostelecky:2008ts}. } independently on the generating model \cite{Abbott:2016blz}.

One important consequence of GW emission by these sources is that one might expect also an EM counterpart to be emitted together with the GW signal depending on the type of the emission source. If an EM counterpart is detected, the time difference observed between the GW and EM signals would be of great importance in many aspects. So far the only source which has been observed in both GWs and EM waves is the GW170817 source \cite{TheLIGOScientific:2017qsa}. On the other hand, for the earlier detected sources of GWs,  the search for EM counterparts by AGILE \cite{Verrecchia:2017lya} and FERMI-LAT \cite{Fermi-GBM:2017orf} collaborations, have reported no significant detection of EM counterpart where and when the event GW170104 was observed by LIGO. Even in the case of the GW150914 event, there has not been any detection of EM counterpart to be directly associated with the GW150914 source \cite{Connaughton:2016umz}.

Although the FERMI-LAT \cite{Fermi-GBM:2017orf} collaboration has not detected any EM counterpart to the GW170104 event, it is quite interesting however to note that AGILE \cite{Verrecchia:2017lya} collaboration has reported a weak EM signal, with signal to noise ratio of $4.2\sigma$, received  $0.46\pm 0.05$ s \emph{before} the detection of the GW170104 event by LIGO. Even though this signal has not been yet associated with the GW170104 event, it is very interesting to investigate if it has been emitted from the same source and if simultaneously with the GW170104 event. On the other hand, in the case of GW170817 event detected by LIGO, the first direct detection of EM counterpart, associated with the GW170817 source, has been detected by FERMI-GMB \cite{Goldstein:2017mmi} roughly 1.7 s \emph{after} the detection of the GW event by LIGO.

The search for EM counterpart, associated with GW emission and their possible simultaneous or time difference detection, is based on the fundamental presupposition of the equality between the speeds of EM and GW signals, namely $v_\text{gw}=v_\text{ph}=1$. Assuming, that both signals are emitted from the same source either simultaneously or their emission time difference is supposed to be known in some way, an important question which naturally arises, if one would detect these signals simultaneously or with the same time emission and detection differences. In another way, supposing that GW and EM signals are emitted simultaneously (time difference equal to zero) from the same source and if GWs travel with speed $v_\text{gw}=1$, is it possible for photons to travel with a speed greater than that of gravitons, namely $v_\text{ph}>v_\text{gw}$? 

The answer to this question essentially depends on the modification of Maxwell equations in order to take into account QED effects in curved spacetime. In fact, as first shown in Ref. \cite{Drummond:1979pp}, by taking into account the one loop vacuum polarization in an external gravitational field, the Maxwell equations in curved spacetime get modified. Depending on the form of the metric tensor $g_{\mu\nu}$, photon superluminal velocities, and gravitational birefringence effects are indeed possible\footnote{Another possibility of faster photons than gravitons has been studied in Ref. \cite{Kostelecky:2015dpa}}.

One main issue related to superluminal signals is the causality structure of spacetime events. However, as shown in Ref.  \cite{Liberati:2001sd}, special relativity safely admits faster than light propagation signals. On the other hand, the question if these signals preserve the causality of events is more subtle and essentially depends on the particular generating mechanism \cite{Liberati:2001sd}. In the case of one loop vacuum polarization in gravitational field \cite{Drummond:1979pp}, some studies show \cite{Khriplovich:1994qj} that the results obtained in Ref. \cite{Drummond:1979pp} are independent on the incident photon wavelength and some others \cite{Shore:1995fz} show that the results  obtained in Ref. \cite{Drummond:1979pp} are valid only at long wavelengths.

\begin{figure}[h!]
\begin{center}
\includegraphics[scale=0.6]{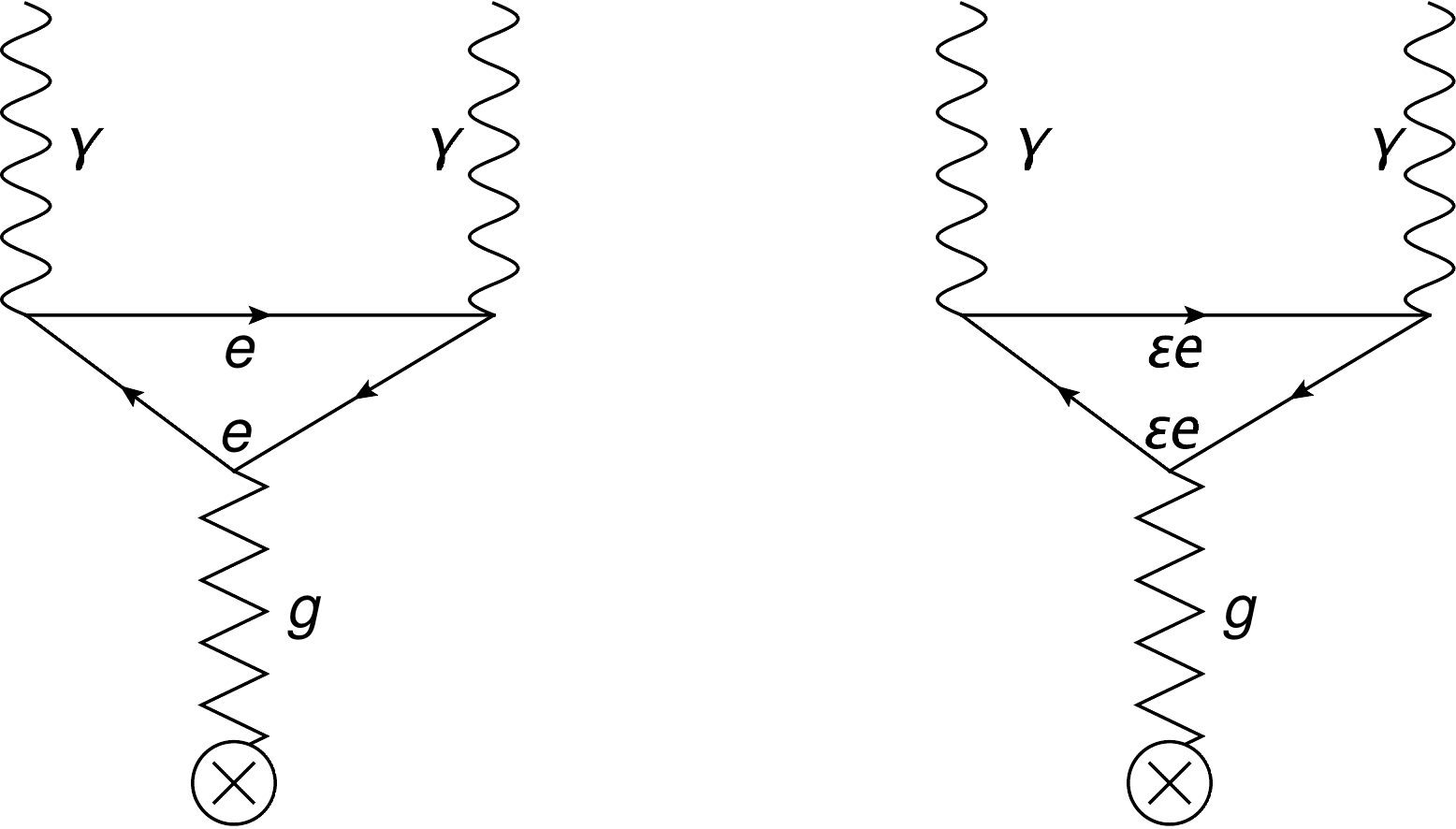}
\caption{Vacuum polarization in external gravitational field. On the left figure the triangle diagram is made of electron-positron pair while on the right figure the triangle diagram is made of millicharged fermion-antifermion pair. }
\label{fig:1}
\end{center}
\end{figure}

The recent observations by LIGO and VIRGO of the GW events and the detection of the EM counterparts give a unique possibility to test if the photon velocity could be superluminal at cosmological scales and if it exceeds the velocity of gravitons. In this paper, I study such possibility and show that millicharged fermions could be masquerading behind the scene by causing superluminal photon velocity. More precisely, I consider the one loop vacuum polarization in an external gravitational field, namely in an expanding universe and study the possibility of superluminal photon propagation. Here, I consider two possibilities of one loop vacuum polarization; the first by considering the loop made of electron-positron pair and the second by considering the loop made of millicharged\footnote{For basic concepts on what are millicharged particles and bounds on the relative charge and mass, see Ref. \cite{Davidson:2000hf}} fermion-antifermion pair (with mass $m_\epsilon$ and charge $Q_\epsilon=\epsilon e$ with $\epsilon$ being the relative charge and $e$ being the electron charge) as shown in Fig. \ref{fig:1}. In addition, I use the velocity of superluminal photons to constrain the parameter space of millicharged fermions.

\section{Vacuum polarization in an expanding universe}
\label{sec:2}

We start with the total action describing the propagation of photons in curved space-time, which is composed of two terms, $S=S_0+S_1$, where $S_0=-(1/4)\int d^4x \sqrt{-g} F_{\mu\nu}F^{\mu\nu}$ is the free photon action minimally coupled to gravity and $S_1$ is the effective action describing one loop vacuum polarization in a background gravitational field \cite{Drummond:1979pp}
\begin{equation}\label{DH-langr}
S_1  =\frac{1}{m_e^2}\int d^4 x\sqrt{-g}\, \left(\mathcal {A} R F_{\mu\nu}F^{\mu\nu} + \mathcal B R_{\mu\nu}F^{\mu\rho}F_{\; \; \rho}^\nu + \mathcal C R_{\mu\nu\rho\sigma} F^{\mu\nu}F^{\rho\sigma} + \mathcal D\, \nabla_\mu F^{\mu\nu}\nabla_\sigma F_{\; \; \nu}^\sigma\right),
\end{equation}
where $m_e$ is the electron mass, $g$ is the metric determinant, $F^{\mu\nu}$ is the total electromagnetic field tensor, $R$ is the Ricci scalar, $R_{\;\;\;\nu\alpha\beta}^\mu$ is the Riemann tensor and $\nabla_\mu$ is the covariant derivative in curved space-time. Here the coefficients $\mathcal {A}, \mathcal B, \mathcal C$ and $\mathcal D$ are respectively given by $\mathcal A=-\alpha/(144\pi), \mathcal B=26\alpha/(720\pi), \mathcal C=-\alpha/(360\pi)$ and $\mathcal D=-\alpha/(30\pi)$ with $\alpha$ being the fine structure constant. By neglecting the last term in the effective action \eqref{DH-langr} which gives a correction to the equation of motion of the order $O(e^4)$, we get 
\begin{align}\label{eq-motion}
\nabla_\mu F^{\mu\nu} -\frac{1}{m_e^2} \nabla_\mu \left[ 4\mathcal A RF^{\mu\nu} +2\mathcal B(R_{\;  \; \rho}^\mu F^{\rho\nu} - R_{\; \; \rho}^\nu F^{\rho\mu})+ 4 \mathcal C R_{\quad \rho\sigma}^{\mu\nu} F^{\rho\sigma}\right] & =  0,\nonumber\\
\nabla_\sigma F_{\mu\nu}+\nabla_\mu F_{\nu\sigma}+\nabla_\nu F_{\sigma\mu} &= 0,
\end{align}
where the second equation in \eqref{eq-motion} is the Bianchi identity for $F_{\mu\nu}$.

The next step is to apply Eqs. \eqref{eq-motion} in the case of Friedmann-Robertson-Walker (FRW) metric with line element $
ds^2=dt^2-a^2(t)\left[\frac{dr^2}{1-\kappa r^2}+r^2 d\Omega^2\right]$,
where $d\Omega^2=d\theta^2+\sin^2(\theta)d\phi^2$. As usual, the constant $\kappa=-1, 0, 1$ is the Gaussian curvature of the space, $a(t)$ is the cosmic scale factor, $r$ is the radial comoving coordinate and $t$ is the cosmological time of a comoving observer. Now by writing $F_{\mu\nu}=f_{\mu\nu}e^{i\Theta(x)}$ in the eikonal approximation with $k_\mu=\nabla_\mu \Theta(x)$ and using it in Eq. \eqref{eq-motion}, one gets \cite{Drummond:1979pp}
\begin{equation}
k^2=11\xi^2 \left(\frac{\dot a^2+ \kappa}{a^2}-\frac{\ddot a}{a}\right)(k\cdot u)^2,
\end{equation}
where $k^2=k_\mu k^\mu$ with $k^\mu=(\omega, \boldsymbol k)$ and $u^\mu=(1, 0, 0, 0)$ is the four-velocity in the rest frame of the medium. The velocity of photons, $v_\text{ph}=\omega/|\boldsymbol k|$, in the rest frame of the medium is given by 
\begin{equation}\label{phase-velo}
|v_\text{ph}|=\left[1- 11\xi^2 \left(\frac{\dot a^2+ \kappa}{a^2}-\frac{\ddot a}{a}\right)\right]^{-1/2},
\end{equation}
where $\omega$ is the total energy of the photon, $\boldsymbol k$ is the photon wave-vector and $\xi^2=\alpha/(90\pi m_e^2)$ for electron-positron vacuum polarization and $\xi^2=(\epsilon/ m_e)^2\alpha/(90\pi)$ for millicharged fermion-antifermion vacuum polarization. 

The expression \eqref{phase-velo} can be further simplified by making use of Friedemann equations, which for the $\Lambda$CDM model read
\begin{equation}\label{F-equations}
\frac{\dot a^2+\kappa}{a^2}=\frac{8\pi G\rho+\Lambda}{3}, \quad \frac{\ddot a}{a}=-\frac{4\pi G}{3}(\rho+3P)+\frac{\Lambda}{3},
\end{equation}
where $P$ and $\rho$ are respectively the pressure and energy density of the (perfect) cosmological fluid and $\Lambda$ is the cosmological constant. Now by using Eqs. \eqref{F-equations} in expression \eqref{phase-velo}, we can write expression \eqref{phase-velo} as
\begin{equation}\label{final-veloc}
|v_\text{ph}|=\left[1- \frac{33}{2}(\xi H_0)^2 \left[\Omega_\text M\left(\frac{a_0}{a}\right)^3+\Omega_\text R\left(\frac{a_0}{a}\right)^4 \right](1+w)\right]^{-1/2}
\end{equation}
where we used the equation of state for the pressure $P=w\rho$ with $w$ being a numerical parameter and $\rho$ being the total energy density of matter and fields only. Here $a_0=a(t_0)$ is the scale factor at present time, $H_0=H(t_0)$ is the Hubble parameter at present time and $\Omega_\text{M}$ and $ \Omega_\text{R}$ are respectively the present time density parameters of matter and radiation. 

We may note that general expression \eqref{final-veloc} for the effective photon velocity does not explicitly depend either on $\kappa$ or $\Lambda$. In addition, it does not depend on the direction of photon propagation namely, it is isotropic in space and it has the same expression for both photon polarization states, namely there is no birefringence. Moreover, since the photon energy $\omega$ is directly proportional to $|\boldsymbol k|$, we have an equality between phase and group velocities of photons.

\section{Superluminal photons and millicharged fermions}
\label{sec:3}

Consider now photons propagating in an expanding universe at post decoupling epoch where the universe is matter dominated with $w\simeq 0$. In this case one can neglect the contribution of relativistic matter to the total energy density and write 
\begin{equation}\label{final-velo}
|v_\text{ph}|\simeq \left\{1- \frac{33}{2}\xi^2 H_0^2 \,\Omega_\text M\left(\frac{a_0}{a}\right)^3\right\}^{-1/2}.
\end{equation}
In order to avoid complex or infinite velocities, we must have that 
\begin{equation}\label{con-1}
\xi <\left[\frac{2}{33\Omega_\text M}\left(\frac{a}{a_0}\right)^3\right]^{1/2}\,H_0^{-1}.
\end{equation}
Suppose that we have a process which emits photons say at initial time $t_i$ where the scale factor value was $a_i=a(t_i)$. In order to have that photon velocities must be real and positive and not formally infinite, we must have that expression \eqref{con-1} must be satisfied for all $t_i\leq t\leq t_0$. Consequently, for fixed values of the parameters $H_0^{-1}$ and $\Omega_\text R$, the condition \eqref{con-1} is satisfied for all $a(t_i)\leq a(t)\leq a(t_0)$ only when $a(t)=a(t_i)$
\begin{equation}\label{con-2}
\xi <\left[\frac{2}{33\Omega_\text M}\left(\frac{a_i}{a_0}\right)^3\right]^{1/2}\,H_0^{-1}.
\end{equation}

So far our presentation has been quite general since we have not yet specified if the vacuum polarization is either due to electron-positron pair or due to millicharged fermion-antifermion pair. In addition, the result given in expression \eqref{con-2} is a constraint on the value of $\xi$ which must be always satisfied. Consider now that vacuum polarization is given by the diagram on the left in Fig. \ref{fig:1}. In this case, $\xi^2=\alpha/(90\pi m_e^2)$ and one can easily verify that for present day values\footnote{In this work cosmological parameters obtained by Planck collaboration are used \cite{Ade:2015xua}. } of $H_0^{-1}\simeq 1.36\times 10^{33}$ eV$^{-1}$ and $\Omega_{\text M}\simeq 0.31$, the condition \eqref{con-2} is always satisfied for values of $a(t)$ at postdecoupling epoch. Moreover, for $\xi^2=\alpha/(90\pi m_e^2)$, one can check from expression \eqref{final-velo} that deviation of the photon velocity from unity is positive and extremely small to have any practical interest for values of $a(t)$ at postdecoupling epoch.

In the case when vacuum polarization is given by the diagram on the right of Fig. \ref{fig:1}, namely due to millicharged fermion-antifermion pair, we have that $\xi^2=(\epsilon/m_\epsilon)^2\alpha/(90\pi)$. In this case the condition \eqref{con-2} becomes
\begin{equation}\label{con-3}
\epsilon/m_\epsilon< 1.18\times 10^{35} (1+z)^{-3/2}\quad  (\text{eV}^{-1}),
\end{equation}
where we used $a_0/a_i=1+z$ with $z$ being the redshift of the source. 

Since we have established that vacuum polarization due to electron-positron pair gives a negligible contribution to the photon velocity at post decoupling epoch, now we focus entirely on vacuum polarization due to millicharged fermion-antifermion pair. It is worth to note that expression \eqref{con-3} has been obtained by simply requiring that the photon velocity in the FRW metric should not be infinite, since for $\epsilon/m_\epsilon\rightarrow 1.18\times 10^{35} (1+z)^{-3/2}\quad  (\text{eV}^{-1})$, the photon velocity given in expression \eqref{final-velo} would be singular.

In order to make our treatment as general as possible, let us assume that we have a source located at a distance $d_s(z)$ which emits GWs and EM waves and these signals are detected either simultaneously or with known time difference. Since both signals travel roughly the same distance\footnote{Here we are neglecting the difference in distance between LIGO and other EM wave detectors such AGILE, FERMI-GMB etc., where the former is located on the Earth while the latter orbits around the Earth. This fact must be properly accounted for in the case of precise experimental tests aiming to measure the simultaneity between the GW and EM signals.} from the source to the detector, we have for the ratio of average velocities\footnote{The photon velocity given in expression \eqref{final-velo} depends on the scale factor and consequently on the cosmological time $t$. This implies that photons travel in an expanding universe with non constant acceleration. } $\bar v_\text{gw}/|\bar v_\text{ph}|=\Delta t_\text{ph}/\Delta t_\text{gw}$ where $\Delta t_\text{gw}$ and $\Delta t_\text{ph}$ are respectively the differences in time between the observed and emitted GW and EM signals in the detector reference frame, namely $\Delta t_\text{gw}=t_\text{gw}^o-t_\text{gw}^i$ and $\Delta t_\text{ph}=t_\text{ph}^o-t_\text{ph}^i$. Here the upper letters $o$ and $i$ indicate respectively the observed and emitted times in the detector reference frame.  Assuming that gravitons travel at the speed $\bar v_\text{gw}=1$, we have that $|\bar v_\text{ph}|=\Delta t_\text{gw}/\Delta t_\text{ph}$. Let be $\Delta t^o=t_\text{gw}^o-t_\text{ph}^o$ and $\Delta t^i=t_\text{gw}^i-t_\text{ph}^i$, then we can write
\begin{equation}\label{time-velo}
|\bar v_\text{ph}|=\frac{t_\text{gw}^o-t_\text{gw}^i}{t_\text{gw}^o-t_\text{ph}^i-\Delta t^o}\simeq 1+\frac{\Delta t^o-\Delta t^i}{t_\text{gw}^o},
\end{equation}
where we used the fact that for cosmological distances, we usually have $t_\text{gw}^o\gg t_\text{ph}^i, \Delta t^o$. 

Since the vacuum polarization in external gravitational field causes superluminal photon propagation, we must have in expression \eqref{time-velo} that $\Delta t^o>\Delta t^i$ if we want that $|\bar v_\text{ph}|>1$. Here $\Delta t^{o, i}$ can be either positive or negative quantities. Next by assuming that deviations from unity of the photon velocities are small, namely for $\epsilon/m_\epsilon \ll  1.18\times 10^{35} (1+z)^{-3/2}\quad  (\text{eV}^{-1})$, we can expand expression \eqref{final-velo} in series at redshift $z$ up to first order and then find the average velocity of photons $|\bar v_\text{ph}|=z^{-1}\int_0^z\,dz^\prime\, |v_\text{ph}(z^\prime)|$ from the emission time at redshift $z$ until today at redshift $z=0$. After by comparing it with expression \eqref{time-velo}, we get the following final expression
 \begin{equation}\label{milli-limit}
\frac{\epsilon}{m_\epsilon}= \left(\frac{1440\pi}{33\,\alpha\,\Omega_\text{M}}\right)^{1/2}\,\left[\frac{\Delta t^o-\Delta t^i}{d_s(z)}\right]^{1/2}\left[\frac{z}{(1+z)^4-1}\right]^{1/2}H_0^{-1}, \quad \text{for}\quad \Delta t^o>\Delta t^i,
\end{equation}
where we used the fact that for GWs, $t_\text{gw}^o=d_s(z)$ for $\bar v_\text{gw}=1$. Expression \eqref{milli-limit} is a consequence of the requirement that photons travel at superluminal velocities in cosmological distances and is valid as far as $\Delta t^o>\Delta t^i$.

By keeping in mind the conditions \eqref{con-3} and \eqref{milli-limit}, let us make as a matter of example some quantitative estimates. Let us consider the case of the GW170817 source detected by LIGO-VIRGO \cite{TheLIGOScientific:2017qsa} with a luminosity distance of $d_L\simeq 40$ Mpc. The corresponding redshift for this source can be found by using numerical integration and we find, $z\simeq 8.9\times 10^{-3}$. The next step is to find the value of $d_s(z)$ for a given source. In an expanding universe dominated by matter and cosmological constant only with $\Omega_\kappa=0$, the distance traveled by GWs (with speed equal to the speed of light in vacuum) emitted by a source located at redshift $z$ is given by
\begin{equation}\label{distance}
d_s(z)\simeq H_0^{-1}\int_0^z\,\frac{dz^\prime}{(1+z^\prime)\,\left[\Omega_\Lambda+\Omega_\text{M}(1+z^\prime)^3\right]^{1/2}}.
\end{equation}
By taking $\Omega_\Lambda\simeq 0.68$ and the redshift of the source GW170817, $z=8.9\times 10^{-3}$, and after integrating \eqref{distance} numerically, we get $d_s\simeq 8.9\times 10^{-3} H_0^{-1}$. By using this value of $d_s$ in expression \eqref{milli-limit}, in Fig. \ref{fig:Fig1a} the plots (solid lines) of $\epsilon/m_\epsilon$ and $m_\epsilon$ as a function of $|\Delta t^i|$ for the observed GW170817 event by LIGO \cite{TheLIGOScientific:2017qsa} and the EM event observed by FERMI-GMB \cite{Goldstein:2017mmi} are shown. For example, the solid line in Fig. \ref{fig:Fig1} represents the values of the ratio $\epsilon/m_\epsilon$, for which the event of GW and EM wave emission at the source, separated by the time difference $\Delta t^i<0$ (namely GW emitted \emph{before} EM waves), is observed at present with the time difference $\Delta t^o=-1.7$ s as found by the observations of LIGO and FERMI-GMB. So, for example with a value of $\epsilon/m_\epsilon\simeq 2\times 10^{29}$ eV$^{-1}$, the GW could be very well emitted $10^4$ s or 2.7 hours before the EM wave at the source and then be detected today 1.7 s before the EM wave.

Another interesting quantity which is important to estimate with given values of $\epsilon/m_\epsilon$ is the distance advance $\delta s$ by superluminal photons that have a positive velocity shift $\delta v$. The physical meaning of $\delta s$ is that it indicates the additional distance traveled by superluminal photons with respect to photons that travel with velocity $|\bar v_\text{ph}|=1$. In the case of vacuum polarization due to the electron-positron pair, one main problem with the distance advance is that it is a very small quantity, namely for the FRW metric one has that $\delta s\ll \lambda_C$ \cite{Drummond:1979pp}. This fact essentially means that is quite difficult to resolve the discrepancy $\delta s$ only with long wavelengths in the case when the derived quantities from the effective Lagrangian density in \eqref{DH-langr} are valid for $\lambda_C <\lambda$. If the derived quantities from the expression \eqref{DH-langr} would also be valid for $\lambda<\lambda_C$, the difficulty on resolving small distance advances would be removed. It is important to note that the results derived in \eqref{milli-limit} and in \eqref{phase-velo} have been found under the additional assumption \cite{Drummond:1979pp} that the photon wavelength must be smaller than the space curvature length scale $L$, namely $\lambda \ll L$. This assumption is valid either the vacuum polarization is due to the electron-positron pair or due to the millicharged fermion-antifermion pair\footnote{The possibility that the distance advance $\delta s$ for QED in curved spacetime is always smaller than $\lambda_C$, independently on the metric, has been discussed in Ref. \cite{Goon:2016une}. In the case of millicharged fermions, we find that the distance advance $\delta s$ calculated by using for example a value of $\epsilon/m_\epsilon\simeq 2\times 10^{29}$ eV$^{-1}$ is smaller than $\tilde\lambda_C=2\pi (2\times 10^{29} \text{eV}^{-1})/\epsilon$ for $\epsilon<1$. Similar results can be found for other allowed values of $\epsilon/m_\epsilon$ in Fig. \ref{fig:Fig1a}.  However, at the current stage we cannot say if this result would hold in general in other situations for millicharged fermions.}.





In the case of millichaged fermions we can calculate the distance advance from the expression \eqref{final-velo} for a source located at redshift $z$ in a universe dominated by matter and vacuum energy only
\begin{equation}\label{unit-dev}
 \delta s=\int_{t_i}^{t_0} \delta|v_\text{ph}|\,dt^\prime =\int_{0}^{z} \frac{\delta|v_\text{ph}|\,dz^\prime}{H_0(1+z^\prime)\sqrt{\Omega_\Lambda+\Omega_\text{M}(1+z^\prime)^3}} \simeq \frac{11}{2}\xi^2 H_0\left[\sqrt{\Omega_\Lambda+\Omega_\text{M}(1+z)^3} - \sqrt{\Omega_\Lambda+\Omega_\text{M}}\right].
\end{equation}
Now by using $\xi^2=(\epsilon/m_\epsilon)^2\alpha/(90\pi)$, we get for the distance advance
\begin{equation}
\delta s\simeq 2.06\times 10^{-42}\left(\frac{\epsilon/m_\epsilon}{\text{eV}^{-1}}\right)^2 \left[\sqrt{\Omega_\Lambda+\Omega_\text{M}(1+z)^3} - \sqrt{\Omega_\Lambda+\Omega_\text{M}}\right] \quad \text{cm}.
\end{equation}
Taking for example a value of $\epsilon/m_\epsilon\simeq 2\times 10^{29}$ eV$^{-1}$ in Fig. \ref{fig:Fig1} which corresponds to $\Delta t^i=-10^4$ s and $z\simeq 8.9\times 10^{-3}$ for the source GW170817, we estimate for the distance advance $\delta s\simeq 3.45\times 10^{14}$ cm. The latter value of the distance advance for millicharged fermions and other values which can be found in a similar way, are several orders of magnitude larger than the typical distance advance which is found in the case of vacuum polarization due to the electron-positron pair, namely $\delta s<\lambda_C$. Consequently, the distance advance generated in the case of vacuum polarization due to millicharged fermions is not a small quantity, on the contrary, it can be comparable to astronomical distances.


\section{Constraints on millicharged fermions}
\label{sec:4}

\begin{figure*}[htbp]
\centering
\mbox{
\subfloat[\label{fig:Fig1}]{\includegraphics[scale=0.55]{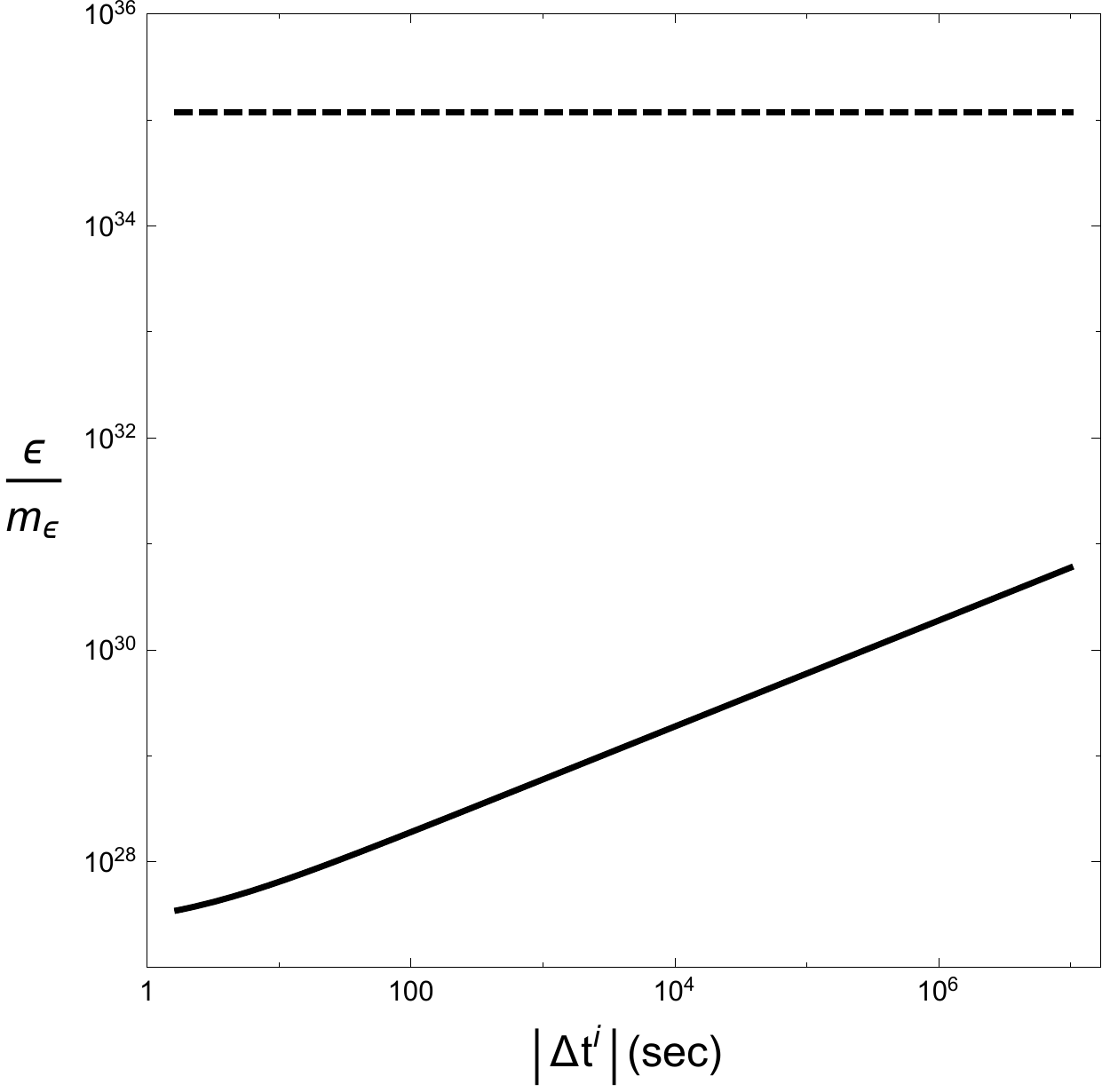}}\qquad
\subfloat[\label{fig:Fig2}]{\includegraphics[scale=0.63]{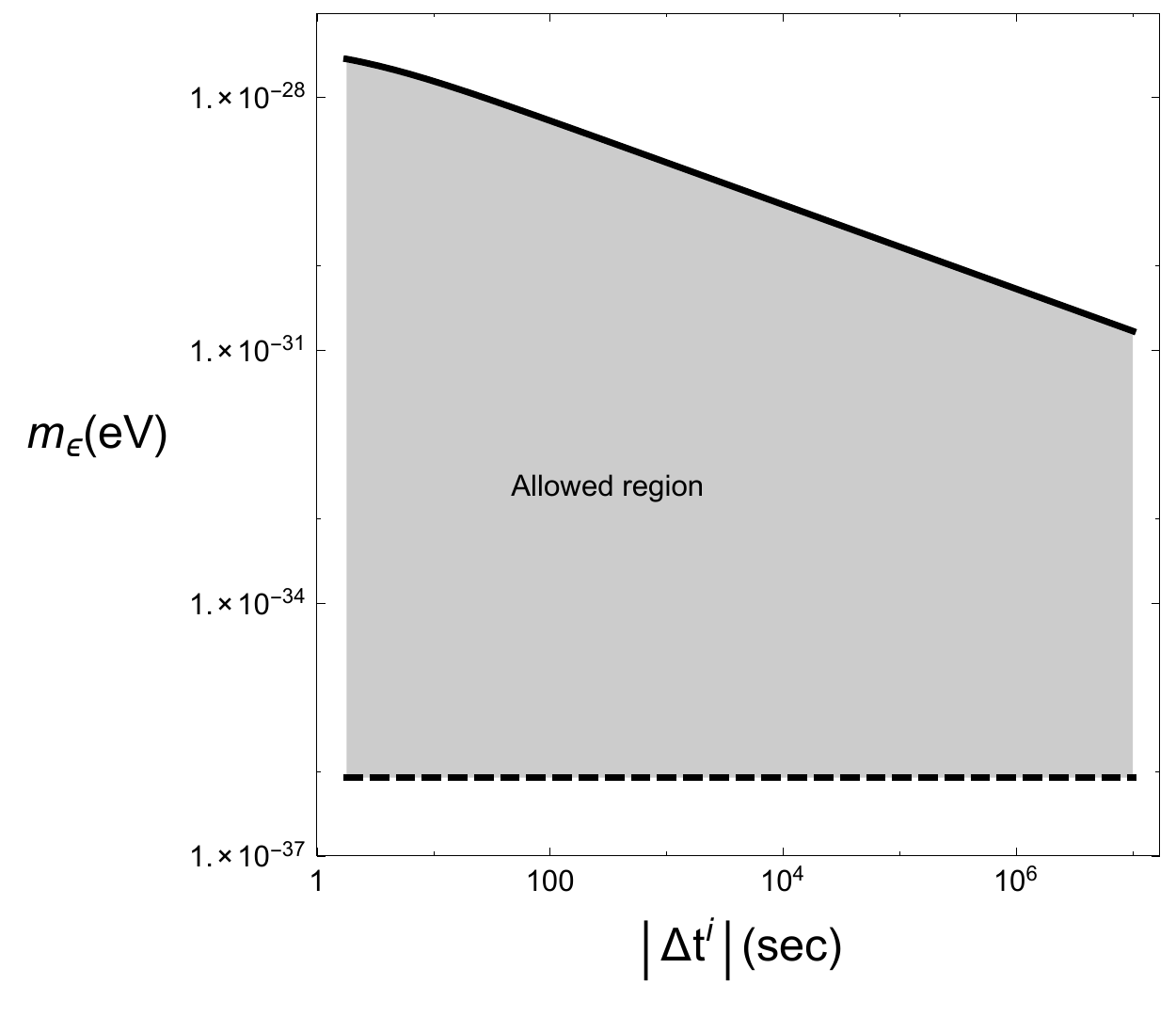}}}
\caption{In (a) the plot of $\epsilon/m_\epsilon$ (in units of eV$^{-1}$) as a function of $|\Delta t^i|$ (the solid line) for $\Delta t^i$ in the interval $\Delta t^i\in [-10^7, -1.8]$ s is shown. The dashed line represent the asymptotic allowed value which the ratio $\epsilon/m_\epsilon$ can have for the millicharged particles that satisfy the condition \eqref{con-3}. In (b) the plot of $m_\epsilon$ as a function of $|\Delta t^i|$ (the solid line) for $\Delta t^i$ in the interval $\Delta t^i\in [-10^7, -1.8]$ s is shown. The dashed line represent the value of the millicharged particle which mass is equal to $m_\epsilon=(33\alpha \Omega_\text{M}/1440\pi)^{1/2}\epsilon\, H_0\left([(1+z)^4-1]/z\right)^{1/2} $ for $\epsilon=0.1$ and $z=8.9\times 10^{-3}$. For smaller values of $\epsilon$ the dashed line shifts toward the bottom of the graphic. The grey region represents the allowed region of values of $m_\epsilon$ as a function of $|\Delta t^i|$ which is obtained by requiring $m_\epsilon\gg (33\alpha \Omega_\text{M}/1440\pi)^{1/2}\epsilon\, H_0\left([(1+z)^4-1]/z\right)^{1/2}$ and by using expression \eqref{milli-limit} for $\epsilon<1$. The solid line is obtained by using the expression \eqref{milli-limit} for $\epsilon=1$. Here we use the value of $\Delta t^o=-1.7$ s which represents the time difference between the observed GW170817 event by LIGO \cite{TheLIGOScientific:2017qsa} and the EM event observed by FERMI-GMB \cite{Goldstein:2017mmi}.}
\label{fig:Fig1a}
\end{figure*}

In Sec. \ref{sec:3} we applied the results found in Sec. \ref{sec:2} to the case of GW170817 event detected by LIGO \cite{TheLIGOScientific:2017qsa} and to the EM event detected by the FERMI-GMB \cite{Goldstein:2017mmi} collaboration. In this specific case we have been able to find values of the ratio $\epsilon/m_\epsilon$, as shown in Fig. \ref{fig:Fig1}, in the case when the time difference between the detected GW and EM waves is $\Delta t^o=-1.7$ s. This fact tells us an important information, namely that limits on the ratio $\epsilon/m_\epsilon$ depend on the time differences $\Delta t^o$ and $\Delta t^i$ (if other parameters are fixed) as is shown in expression \eqref{milli-limit}.

In order to compare our results with existing limits on the parameter space of millicharged particles $\epsilon$ and $m_\epsilon$, is important first to say few words on the approximations used so far. As already mention in Sec. \ref{sec:3}, the derived result for the photon velocity \eqref{final-velo} are valid also for $\lambda\ll L$. Another additional condition comes from the requirement that deviation from unity of the photon average velocity must be small, namely $\delta |\bar v_\text{ph}|\ll 1$ which can be also written as
 \begin{equation}\label{milli-limit-1}
\tilde\lambda_C \ll \left(\frac{5760\pi^3}{33\,\alpha\,\Omega_\text{M}}\right)^{1/2}\,\left(\frac{H_0^{-1}}{\epsilon}\right)\left[\frac{z}{(1+z)^4-1}\right]^{1/2} \quad \text{or}\quad m_\epsilon \gg \left(\frac{33\,\alpha\,\Omega_\text{M}}{1440 \pi}\right)^{1/2}\,\epsilon H_0\,\left[\frac{(1+z)^4-1}{z}\right]^{1/2}
\end{equation}
where $\tilde\lambda_C=2\pi/m_\epsilon$ is the Compton wavelength of the millicharged fermion. As we can see from \eqref{milli-limit-1}, the Compton wavelength of the millicharged fermion, apart from a constant factor, is directly proportional to $H_0^{-1}$ and inversely proportional to $\epsilon$. This fact tells us that while it is possible for $\tilde \lambda_C\geq H_0^{-1}$, we have for values of $\epsilon<1$ that the condition \eqref{milli-limit-1} can be still satisfied.

As it should be clear by now, our result \eqref{milli-limit} gives only the value of the ratio $\epsilon/m_\epsilon$ (for fixed values of other parameters) and it does not allows us to say much on the parameter space of millicharged fermions. The only thing that we know, as far as the concept of millicharged fermions applies, is that $\epsilon<1$. Therefore, if we want to limit or constraint the parameter space of millicharged fermions is necessary to use complementary information on millicharged parameter space obtained either experimentally or by other models. From the experimental side, the PVLAS experiment gives the most tight constraints on the parameter space of millicharged fermions for masses below the eV, namely $\epsilon<3\times 10^{-8}$ for masses $m_\epsilon\lesssim 0.01$ eV, see Fig. 15 of Ref. \cite{DellaValle:2015xxa}. On the other hand, several phenomenological models give different constraints and for a review see Ref. \cite{Davidson:2000hf}. As studied in Ref. \cite{Davidson:2000hf} and shown in their Fig. 1, the model depended results exclude any millicharged fermion in the parameter space $10^{-15}\leq \epsilon< 1$ in the mass range $m_\epsilon\lesssim m_e$. On the other hand for masses $m_e\lesssim m_\epsilon$, the constraint on $\epsilon$ are less stringent in comparison with millicharged fermions with masses $m_\epsilon\lesssim m_e$. These constraints are obtained from a combination of several models that include big bang nucleosynthesis (BBN) constraints, constraints from emission of millicharged fermions by SN1987, constraints from decay of plasmons into millicharged fermions in white dwarfs and in red giants, constraints from invisible decays of orthopositronium etc.

Since we can not constrain $\epsilon$ or $m_\epsilon$ separately, we can use the constraints found by PVLAS or those found in Ref. \cite{Davidson:2000hf} for $\epsilon$ in order to constrain $m_\epsilon$. For example, consider the constraint found by PVLAS on $\epsilon<3\times 10^{-8}$ for $m_{\epsilon}\lesssim 0.01$ eV. For such value of $\epsilon$ we find a much tighter constraint on $m_\epsilon$
\begin{equation}\label{PVLAS}
m_\epsilon < 1.6\times 10^{-35} \left(\frac{f(z)}{\Delta t^o-\Delta t^i}\right)^{1/2}\left[\frac{(1+z)^4-1}{z}\right]^{1/2}\quad \text{eV} \quad (\text{PVLAS})
\end{equation}
where $f(z)$ is defined as
\begin{equation}
f(z) \equiv \int_0^z\,\frac{dz^\prime}{(1+z^\prime)\,\left[\Omega_\Lambda+\Omega_\text{M}(1+z^\prime)^3\right]^{1/2}},
\end{equation}
where $\Delta t^o, \Delta t^i$ are expressed in units of seconds. In the case when $\epsilon<10^{-15}$ for the model dependent constraints, we find 
\begin{equation} \label{MD}
m_\epsilon < 5.33\times 10^{-43} \left(\frac{f(z)}{\Delta t^o-\Delta t^i}\right)^{1/2}\left[\frac{(1+z)^4-1}{z}\right]^{1/2}\quad \text{eV} \quad (\text{Model Dependent}).
\end{equation}

Of course in both expressions \eqref{PVLAS} and \eqref{MD} we must have that $\Delta t^o> \Delta t^i$ as far as the concept of superluminal photon propagation is concerned. If we consider for example $\Delta t^o=-1.7$ sec and $f(z)=8.9\times 10^{-3}$ for the GW170817 source with redshift $z\simeq 8.9\times 10^{-3}$, we get the constraints $\epsilon <3\times 10^{-8}$ and $m_\epsilon< 1.5\times 10^{-36}\left[-1.7-(\Delta t^i/\text{sec})\right]^{-1/2}$ eV for the PVLAS constraint on $\epsilon$. On the other hand for the model dependent limits we get the constraints $\epsilon <10^{-15}$ and $m_\epsilon< 5\times 10^{-44}\left[-1.7-(\Delta t^i/\text{sec})\right]^{-1/2}$ eV. These considerations tell us that experimental and/or model dependent limits on $\epsilon$ give very stringent limits on $m_\epsilon$. The constraints on the masses of these millicharged fermions depend on $\Delta t^i$ and imply that $m_\epsilon$ must be very small, namely ultralight millicharged fermions. On the other hand, independently on the experimental or model dependent constraints on $\epsilon$, the simple fact that for millicharged particles we must have $\epsilon<1$, gives us the constraint on the mass
\begin{equation} \label{MDI}
m_\epsilon < 5.33\times 10^{-28} \left(\frac{f(z)}{\Delta t^o-\Delta t^i}\right)^{1/2}\left[\frac{(1+z)^4-1}{z}\right]^{1/2} \quad \text{eV}.
\end{equation}
The limit \eqref{MDI} would immediately exclude any high mass millicharged fermion, as for example those with masses $m_\epsilon> m_e$, if the difference $\Delta t^o-\Delta t^i$ is not extremely small.

\begin{figure*}[htbp]
\centering
\mbox{
\subfloat[\label{fig:Fig3}]{\includegraphics[scale=0.6]{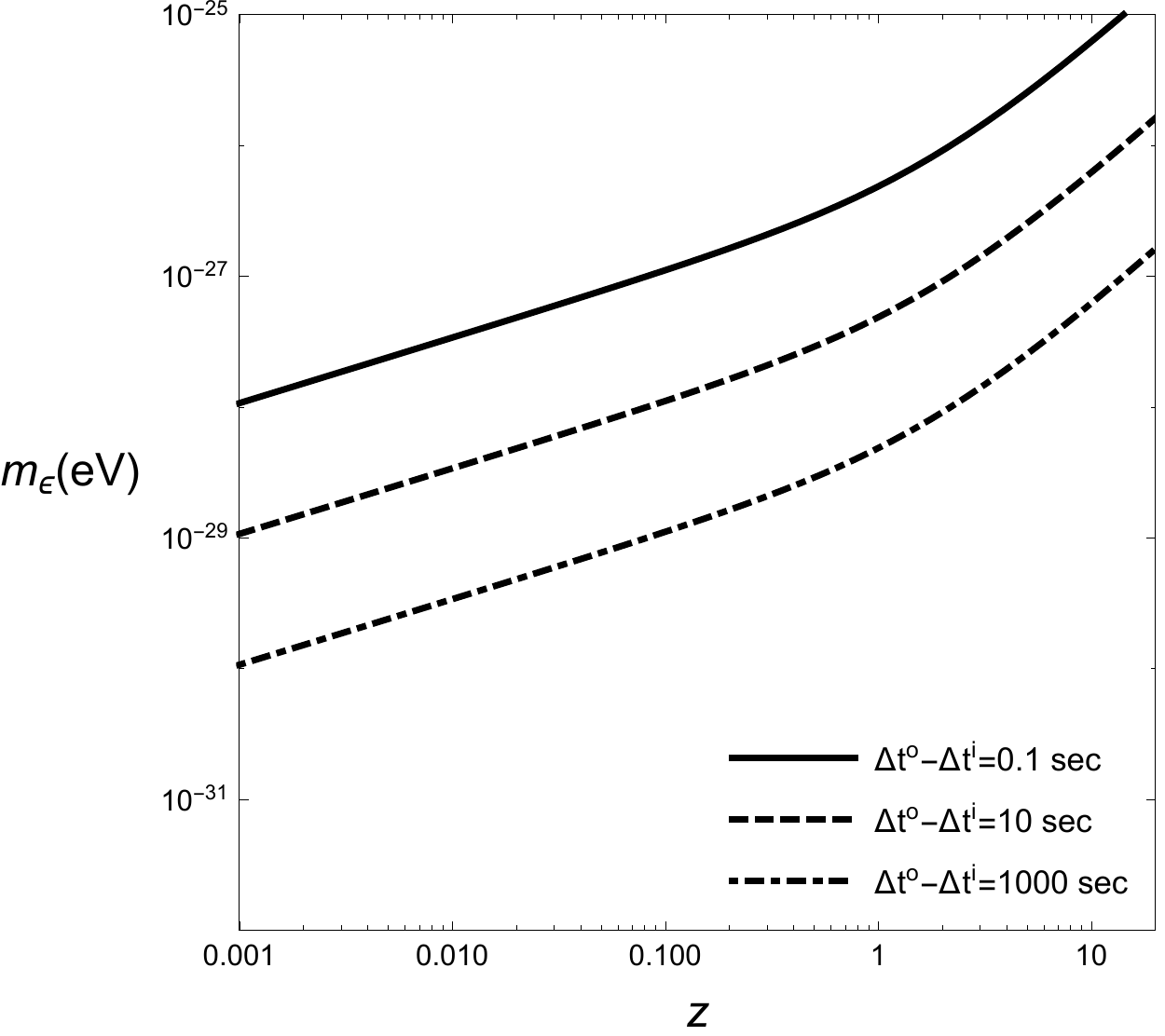}}\qquad
\subfloat[\label{fig:Fig4}]{\includegraphics[scale=0.6]{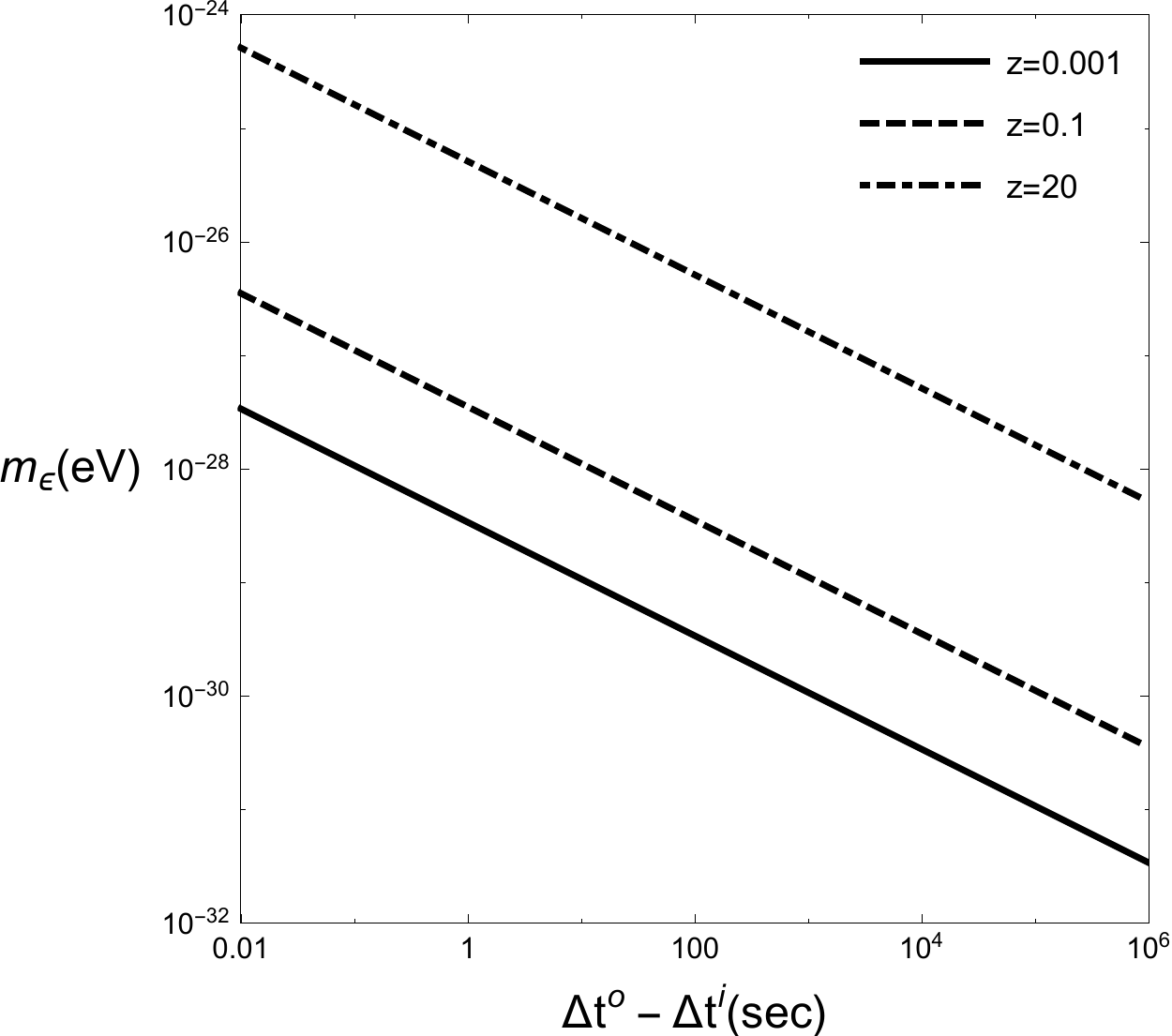}}}
\caption{In (a) the plots of $m_\epsilon$ as a function of the emitting source redshift $z\in [0.001, 20]$ for various values of $\Delta t^o-\Delta t^i$ are shown. In (b) the plots of $m_\epsilon$ as a function of $\Delta t^o-\Delta t^i$ in the interval $(\Delta t^o-\Delta t^i)\in [0.01, 10^6]$ sec for various values of the source redshift are shown. In both figures the plots have been made by using the constraint \eqref{MDI} for $\epsilon<1$. The region above each line is excluded while the region below each line is allowed. As we can see in (a), the smaller is the difference in $\Delta t^o-\Delta t^i$, the less stringent are the constraint on $m_\epsilon$. In (b) the smaller is the redshift of the emitting source, the more stringent are the constraints on $m_\epsilon$.}
\label{fig:Fig2a}
\end{figure*}

\section{Conclusions}
\label{sec:5}

We have shown that in the case when one takes into account the vacuum polarization due to millicharged fermions, the propagation velocity of photons in an expanding universe exceeds that of gravitational waves depending on the ratio $\epsilon/m_\epsilon$. These results have been obtained in the case when the derived quantities from the expression of the effective action \eqref{DH-langr} would be valid for those photon wavelengths with $\lambda\ll L$, where $L\geq H_0^{-1}$ for a spatially flat universe up to the Hubble distance,  and independently if $\lambda$ is either smaller or bigger with respect to $\tilde\lambda_C$ \cite{Khriplovich:1994qj}. If the derived quantities from the effective action \eqref{DH-langr}, correctly describes vacuum polarization in gravitaional field only for $\lambda>\lambda_C$ or $\lambda>\tilde\lambda_C$, which in the case of vacuum polarization due to millicharged fermions translates\footnote{Since the deviation from unity of the index of refraction is proportional to $\xi^2$ and very small for millicharged particles, one can still use the relation $\nu=|v_\text{ph}|/ \lambda \simeq 1/\lambda$ for photons.} to $\nu<\nu_C(m_\epsilon/m_e)$ (where $\nu_C=8.16\times 10^{-2}$ MeV is the Compton frequency of the electron), then one should treat with care the correspondence between the observation frequency of EM detectors and the limit of validity of the theory. Nevertheless, the result obtained in \eqref{con-3} is valid even when $\nu<\nu_C(m_\epsilon/m_e)$. 

Our result in \eqref{milli-limit} and the plots in Fig. \ref{fig:Fig1a} shown for the case of GW170817 event detected by LIGO and for the EM event detected by FERMI-GMB, indicate that millicharged fermion vacuum polarization can cause photon superluminal velocities in expanding universe depending on the ratio $\epsilon/m_\epsilon$. The superluminal photon velocity in expanding universe implies that if a source emits both GWs and EM waves with time differences satisfying the condition $\Delta t^o>\Delta t^i$, the EM signal could be observed either in advance or later with respect to the GW signal. In the case when $\Delta t^i \geq 0$, the EM signal can be observed only in advance. In the case when $\Delta t^i<0$ (the GW emitted before the EM wave at the source), the EM signal can be observed simultaneously if $\Delta t^o=0$ or in delay or in advance with respect to the GW signal depending on the sign of $\Delta t^o$ with respect to $\Delta t^i$ and on their respective magnitudes.

As shown in Figs. \ref{fig:Fig1a} and \ref{fig:Fig2a}, the masses of millicharged fermions that causes photon superluminal propagation in expanding universe are quite small (if the difference $\Delta t^o-\Delta t^i$ is not extremely small), namely ultralight millicharged fermions.
An interesting fact and at the same time a disturbing one is that the photon velocity for given values of the ratio $\epsilon/m_\epsilon$ can be very large. For example, in the case of the GW170817 event detected by LIGO and for the EM event detected by FERMI-GMB, the observed time difference of $\Delta t^o=-1.7$ s does not necessarily mean that these signals were emitted with the same initial time difference. It could be very well, depending on the value of $\epsilon/m_\epsilon$, that the GW signal could have been emitted few seconds or hours or even years before the EM signal and then be detected today with a time shift of $\Delta t^o=-1.7$ s with respect to the EM signal. This fact would create serious difficulties in establishing how a source which emits EM waves and GWs evolved in time and when various processes at the source occurred.

\vspace{+1.5cm}

 {\bf{AKNOWLEDGMENTS}}:
 This work is supported by the Grant of the President of the Russian Federation for the leading scientific Schools of the Russian Federation, NSh-9022-2016.2.

\end{document}